\documentclass[usletter, 10pt, conference]{ieeeconf}      
                                                          
\IEEEoverridecommandlockouts                              
\usepackage{graphicx} 
\usepackage{epsfig} 
\usepackage{amsmath} 
\usepackage{amssymb}  

\graphicspath{{img/pdf/}{img/png/}}
\usepackage{booktabs} 
\usepackage{capt-of}
\usepackage{mathtools}

\usepackage{url}
\usepackage{algorithm} 
\usepackage{algorithmicx} 
\usepackage{algpseudocode} 
\usepackage[dvipsnames,table,xcdraw]{xcolor} 
\usepackage{mathtools} 
\usepackage{bm}
\usepackage{color}

\newcommand{\A}{\texttt A}
\newcommand{\C}{\texttt C}
\newcommand{\G}{\texttt G}
\newcommand{\T}{\texttt T}

\title{\LARGE \bf
Sequencing on Silicon: AI SoC Design for Mobile Genomics at the Edge
}

\author{Sebastian Magierowski$^{1}$, Zhongpan Wu$^{1}$, Abel Beyene$^{1}$, Karim Hammad$^{2}$
\thanks{*This work was supported by NSERC, Qualcomm, and Globalfoundries.}
\thanks{$^{1}$Lassonde School of Engineering, Electrical Engineering and Computer Science, York University, Toronto, Canada (\texttt{magiero@eecs.yorku.ca})}
\thanks{$^{2}$Arab Academy for Science, Technology, and Maritime Transport, Cairo, Egypt.}
}

\begin{document}
\maketitle
\thispagestyle{empty}
\pagestyle{empty}

\begin{abstract}
Miniature DNA sequencing hardware has begun to succeed in mobile contexts, driving demand for efficient machine learning at the edge. This domain leverages deep learning techniques familiar from speech and time-series analysis for both low-level signal processing and high-level genomic interpretation. Unlike audio, however, nanopore sequencing presents raw data rates over 100× higher, requiring more aggressive compute and memory handling. In this paper, we present a CMOS system-on-chip (SoC) designed for mobile genetic analysis. Our approach combines a multi-core RISC-V processor with tightly coupled accelerators for deep learning and bioinformatics. A hardware/software co-design strategy enables energy-efficient operation across a heterogeneous compute fabric, targeting real-time, on-device genome analysis. This work exemplifies the integration of deep learning, edge computing, and domain-specific hardware to advance next-generation mobile genomics.
\end{abstract}

\section{Introduction}

For over 150 years biology has been a leader among the sciences in the application of statistical data analysis methods, a reflection of the `data rich, but model poor' nature of the field and it's immense importance to our self-understanding, survival, and flourishing.  This leadership has carried through to the current swell in machine learning (ML) methods and their extension to artificial intelligence (AI).  In particular, over the last 40 years, molecular biology has benefitted from from early connectionist advances~\cite{Sejnowski88} all the way through recent~\cite{AlphaFold23} advances in deep learning (DL).  

For electronic hardware engineers focused on applications at the edge, an exciting frontier is being opened by this field.  Already, instruments once confined to experimental physiology are commonly packaged into wearables with associated low-power computing elements.  The same is increasingly occurring with other life science staples like microscopy (enhanced smartphone imaging of tissue)~\cite{Song25smartphoneimage}, biochemistry (mini spectrometry for material composition)~\cite{vanKollenburg21Scio}, molecular biology (mini molecular amplification for pathogen detection in plants)~\cite{Yadav25Biomeme}, etc.  All of these present valuable opportunities for embedded AI to locally process and interpret the biomeasurements.

\begin{figure}
\centerline{\includegraphics[width=3.5in]{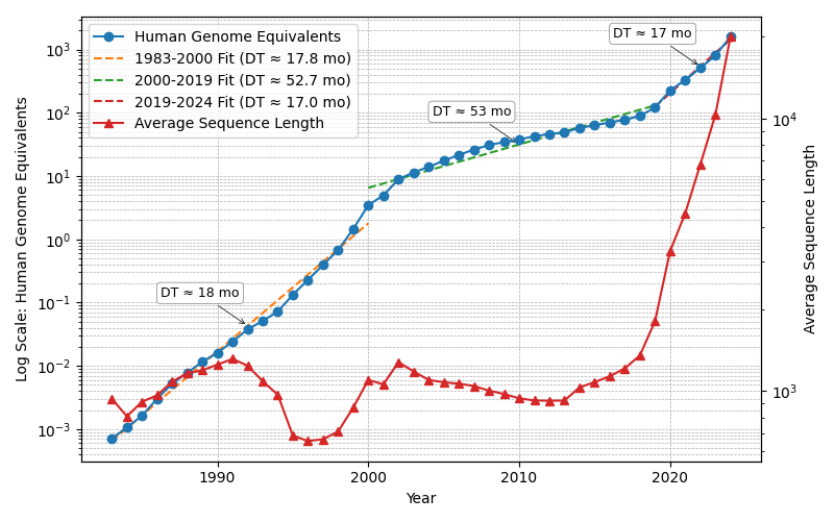}}
\caption{The growth of genomic molecular data in GenBank over time.}
\label{f:genbank}
\end{figure}

In this paper we discuss the design opportunities and implications of such bioinstrumentation+AI work in the context of another biological discipline: genomics.  Genomics is among the leaders in scientific data collection with just one database, the Sequence Read Archive (SRA) currently holding 47 PB of publicly available data~\cite{Sayers25SRA}, for comparison, CERN's open data portal grants access to about 5 PB~\cite{CERNOpenData25}).  The growth rate of post-processed genomic data as exemplified by the assembled sequences in GenBank, another famous genomics database, is also impressive, as shown in Fig.~\ref{f:genbank}.  This data reflects not only increased measurement over time, but increased effort to compute on this data.  The new growth rate records, a database size doubling time (DT) of 17 months over the last five years, shows a clear spurt in activity as analysis methods mature and as more ambitious genome projects are pursued.  Also, the rapid growth in average GenBank sequence size, a 20${\times}$ jump in five years, reflects important development of relevance to bioinformatic AI in edge devices.

\begin{figure}
\centerline{\includegraphics[width=3.4in]{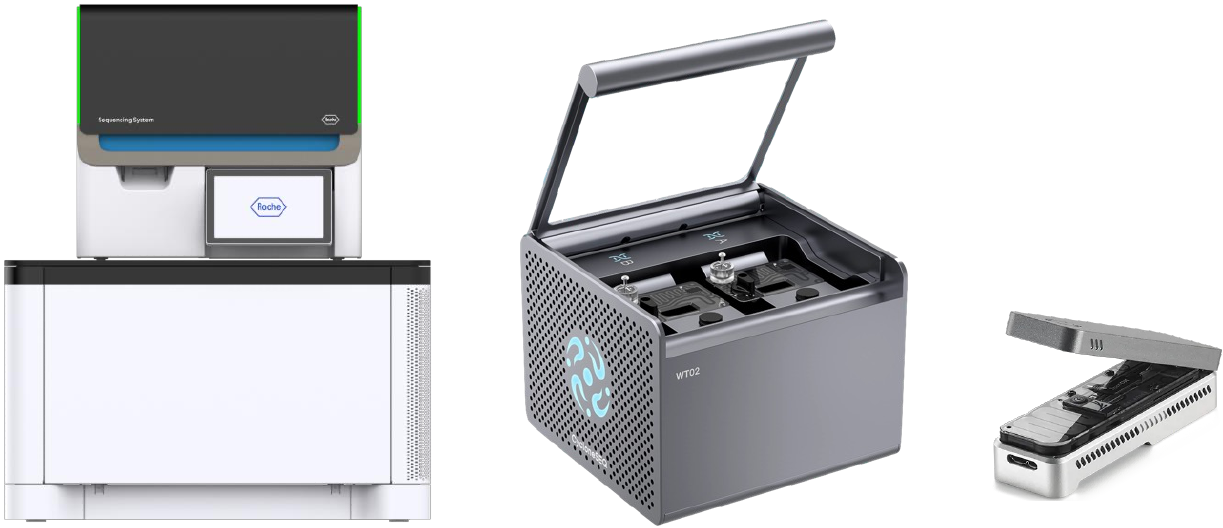}}
\caption{Examples of recent miniature real-time DNA sequencing machines.}
\label{f:miniseqs}
\end{figure}

Like the examples given above, genomics too is benefitting from the emergence of smaller front-end equipment in the form of miniature DNA sequencers (examples are shown in Fig.~\ref{f:miniseqs}).  Besides their small size, these devices can also measure strands more than 100${\times}$ longer than large machines, presenting new measurement analysis challenges for computing.  Even more importantly, these sequencers produce raw measurement results in real-time, presenting workloads for bioinformatic AI not unlike those tasked with real-time video, sound, and language processing in commodity smartphone devices.  The fact that these sequencers now incorporate CMOS as integral parts of their function also opens the door for especially effective integration of edge AI solutions.

The remainder of the paper is organized as follows.  In \S~II we discuss the genomic workloads to which edge processing may need to be applied are outlined.  As part of this discussion we define and contextualize our view of edge hardware and then describe genomics workloads with reference to this.  Key CMOS AI contributions in this space are also highlighted. In \S~III we describe a SoC designed facilitate AI computation for genomics edge uses and give preliminary results. \S~IV concludes the paper with a summary discussion.

\section{Genomic Workloads at the Edge}

\subsection{Opportunities Across the RAN}

In assessing genomics at the edge, we take a broader view and consider the possibility of hosting applications on either side of the radio access network (RAN).  That is, we imagine opportunities at the multi-access edge computing (MEC) nodes~\cite{Cruz22edge}, through personal mobile devices~\cite{Murshed21mledge} to tinyML embedded units~\cite{Abadade23tinyML}.  This reflects the current partition of the MLCommons (MLC) hardware benchmark suite that, among other categories, assesses DL inference performance at ``Edge'', ``Mobile'', and ``Tiny'' tiers.  Table~\ref{t:mlc} summarizes the types of benchmark models, hardware, and performance aspects in these categories.  Therein, “Tiny” refers to microcontroller-class or ultra-low-power chips (tens of MHz, kilobytes to a few MB of memory). “Mobile” covers smartphone-class devices (multi-core ARM CPUs, mobile NPUs/GPUs, a few GB of RAM). “Edge” includes more powerful embedded hardware like single-board computers, Jetson-class devices, FPGAs, or on-instrument accelerators (often tens of watts power and 8–16GB RAM).  And although MEC devices are clearly not mobile themselves, their presence can be an indespensible facilitator of mobile AI and genomics applications.

\newcolumntype{L}[1]{>{\raggedright\arraybackslash}p{#1}}
\renewcommand{\arraystretch}{1.4}
\begin{table}
\centering
\scriptsize
\caption{Hardware and ML across the RAN: Edge vs. Mobile vs. Tiny}
\begin{tabular}{L{1.1cm} L{1.9cm} L{1.9cm} L{1.9cm}}
\toprule
\textbf{} & \multicolumn{1}{c}{(Multi-Access)~\textbf{Edge}} & \multicolumn{1}{c}{\textbf{Mobile} (Edge)}  & \multicolumn{1}{c}{\textbf{Tiny} (Edge)} \\
\midrule
\textbf{Model Size}         & 50M–6B+ param.        & 2M–25M param.                & $<$1MB mem.\\
\textbf{Latency}            & real-time \& batch    & real-time                    & low latency \\
\textbf{Power}              & 1–10W                 & 0.1–1W                       & $<$10mW \\
\textbf{Hardware}           & small server (GPU, NPU, TPU) & smartphone (NPU, SoC) & microcontroller (DSP, SoC) \\
\textbf{Examples}           & Jetson                & Snapdragon                   & Cortex-M \\
\textbf{Deployment Context} & on-premises           & personal, wearable, cameras  & embedded sensors, toys, IoT \\
\textbf{Network Types}      & GPT-J, 3D-UNet, BERT, SDXL, RNNT & MobileNet, Mobile-BERT, Stable Diffusion & DS-CNN, ResNet-8, Autoencoders \\
\textbf{Test Datasets}      & ImageNet, COCO, SQuAD & ImageNet, ADE20K, COCO       & CIFAR-10, Speech Commands \\
\bottomrule
\end{tabular}
\label{t:mlc}
\end{table}

Roughly, as per MLC's current outlook, MEC nodes are benchmarked on 50M-6B parameter DL loads with tasks spanning language and speech processing and vision.  This also includes important biological applications like medical image segmentation.  As noted, smartphones and embedded devices are representative hardware for the mobile and tiny categories, respectively.  In the mobile case, networks below about 25M parameters are benchmarked as they might be used to satisfy use-cases spanning content creation, augmented reality and smart camera assistance, voice assistance and real-time translation, etc.  In the tiny ML case, very small models, often constrained to less than a 1-MB memory footprint, are run on small chips for applications like wake-word detection, small drones and automobile sensors, anomaly detection in bio or industrial signals, etc.

\subsection{Genomics Workload Pipelines}

\begin{figure}
\centerline{\includegraphics[width=3.5in]{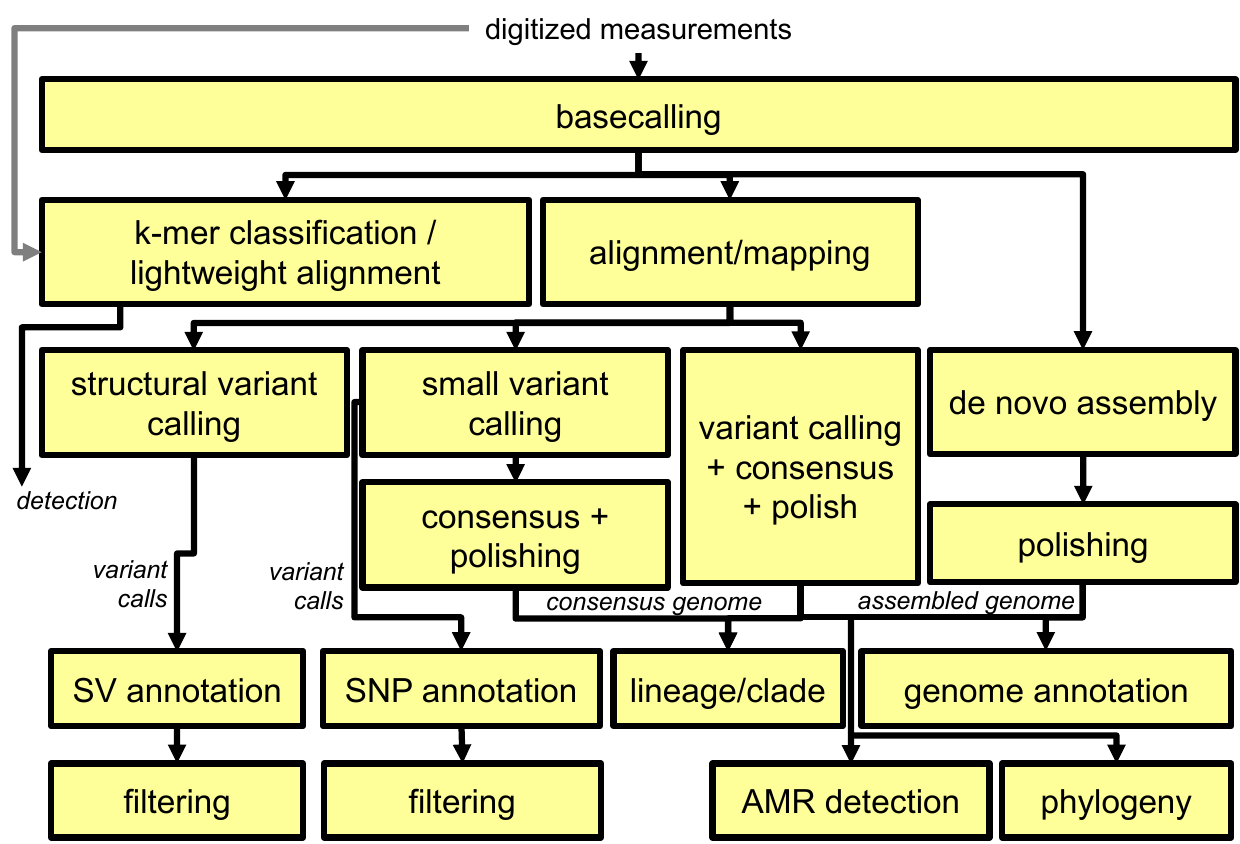}}
\caption{A summary sketch of sequencing computational superstructure.}
\label{f:seqsuper}
\end{figure}

How can genomics workloads reasonably map onto the edge-mobile-tiny axis?  This is a complex question as the workload is intricate and can take on a multitude of computational ``sequencing pipelines'' depending on the answers one seeks.  A summary of computational superstructure from which these pipelines may be extracted is shown in Fig.~\ref{f:seqsuper}.  Naturally, the large majority of these steps involve some kind of biological sequence analysis, which has been strongly influenced by linguistics~\cite{Baldi01}.  We highlight some of the steps in Fig.~\ref{f:seqsuper} now.

\subsubsection{Basecalling}
An unambiguously critical stage, because it usually precedes all ensuing pipeline computations, is {\it basecalling}.  This step is particularly vital in miniature portable long-read devices (Fig.~\ref{f:miniseqs}).  Just as automatic speech recognition (ASR) systems convert raw audio waveforms into text transcripts, basecallers transform digitized electrochemical signals from DNA sensors~\cite{Dawji21} into nucleotide sequences — effectively serving as genomic ASRs.  These output ``base'' sequences are called {\it reads}, strings of text composed of the letters \A,\C,\G,\T.  For large sequencing machines, millions of reads, each consisting of 1000s of bases can be output by a basecaller per hour.  And as with ASRs basecalling, at least in the context of miniaturizable, long-read, sequencers has come to be dominated by DL methods.  Also, from a networking perspectives, basecallers present the opportunity to effectively compress high bandwidth measurements into low-bandwidth text representations, a valuable benefit in wireless settings.  For example, whereas mono voice data might represent a flow of raw data on the order of 256~kbps, hand-sized DNA sequencers, not counting meta-data needs, can easily exceed this by 100${\times}$ and reach 30~Mbps of real-time sensory data throughput.

What scale of problem does basecalling represent?  The numbers can vary wildly.  For very precise applications $\sim$50 GFLOP/sec/DNA sensor are needed~\cite{AmazonBasecall23}.  Keeping in mind that about 1000 sensors fit in a thumbnail sized-footprint it is not surprising that for large sequencers even powerful banks of GPU servers take many hours to complete this task.  As the technology improves, the DNA sensor density will surely increase.  For reference, a simple 39M parameter ASR model~\cite{Radford23whisper} requires roughly $\sim$0.7~GFLOP/sec.

As genomics tasks vary and the need for accuracy drops, it becomes possible to simplify models~\cite{Cavlak24targetcall}.  Thus, models needing as little as $\sim$60~MFLOP/sec/sensor may be reasonable.  But given that in aggregate, we are possibly dealing with 1000s of sensors in aggregate, even modest basecalling algorithms may quickly require the resources of advanced smartphones.

To our knowledge, only one fabricated basecalling ASIC has been produced to date~\cite{Dawji23}, a RISC-V based SoC with accelerated Viterbi processing.  In a 22-nm CMOS technology this device was able to process about 30\,Kbase per second within about 20\,mW at 200\,MHz, a 200${\times}$ energy efficiency improvement over a 16-nm ARM Cortex-A53 SoC.  These numbers highlight the potential of porting this function even down to the Tiny hardware tier.

\subsubsection{Alignment}
This is arguably the most intensely researched part of the sequencing pipeline: the process of matching reads to a pre-built reference genome. For example, a recent survey catalogs 107 sequence alignment methods introduced over the past 30 years~\cite{Alser21align}. A dominant paradigm for alignment is the seed-and-extend method. The seed step, based on a contextualized reorganization of the reference genome (the Burrows–Wheeler Transform) and its efficient indexing (FM-index), allows rapid search for very short exact matches (typically $\sim$10\,bases). The following step, extension, vets promising seeds by computing an approximate dynamic programming (DP) alignment. Although neither seeding nor extension employs deep learning (DL), DP itself has long been the computational substrate for many ML workloads. Thus, in the context of Sutton's Bitter Lesson~\cite{Sutton19}, DP—like DL—represents a generalizable algorithmic structure that favours scalable, hardware-accelerated implementation over hand-crafted heuristics. Moreover, efficient DP implementations can be reused across a range of similar tasks throughout the sequencing pipeline.

Perhaps even more surprisingly than basecalling, very little in the way of customized ASIC has been experimentally demonstrated in this space.  A standout example for DP-based extension is~\cite{Wang21} in which a 8-mm$^2$, 55-nm CMOS design is described capable of extending the equivalent of about 0.25\,Gbases per second for about 500\,mW at 670\,MHz.  This is about a 70${\times}$ energy efficiency improvement over FPGA implementations and seems like a promising implementation in the Mobile tier.

\subsubsection{Variant Calling}
Atop, the ``primary analysis'' pillars of basecalling and alignment, are a vast set of bioinformatics steps that extract ever-more actionable information from the genomic data.  These are the ``secondary analysis'' and ``tertiary analysis'' phases of the possible sequencing pipelines.  Among the most outstanding examples here is variant calling, the process of identifying key differences between the genome of the measured organism and that of a representative (reference) genome.  Here, FPGA-accelerated systems can align and variant call a clinically relevant human genome in about 30\,min for about 50\,W~\cite{Dragen25} a job on the upper range of the Edge tier identified above.

Aspects of these workloads are now very well entrenched in the DL domain, with tools like DeepVariant and Clair3 being excellent examples of such~\cite{Abdelwahab25}.  Although DeepVariant represents a heavier workload (an Inception-based model with $\sim$25M parameters) more compact systems lighter systems like Clair complete clinical-level human genome variant calls in about 5 hours over 24 threads without GPU support~\cite{Luo20}.  Even more promising results have been reported with a report showing that the SARS-Cov-2 genome of $\sim$30\,Kbases could be analyzed within about 30 min on a smartphone~\cite{Genopo20} thus indicating the potential for in-depth viral analysis at the Mobile tier.

In terms of experimental CMOS ASIC results, this area also remains sparse with the most recent notable contribution being a 16-mm$^2$, 28-nm CMOS chip capable of mapping, variant calling, and genotyping at 95\,Mbases per second for about 2{,}700 mW at 400-MHz~\cite{Wu25}, a system possibly realistic for to the Mobile tier if advanced nodes in the 5 to 3-nm range are considered.

\section{A Proposed SoC For Mobile Genomics}

\begin{figure}
\centerline{\includegraphics[width=2.4in]{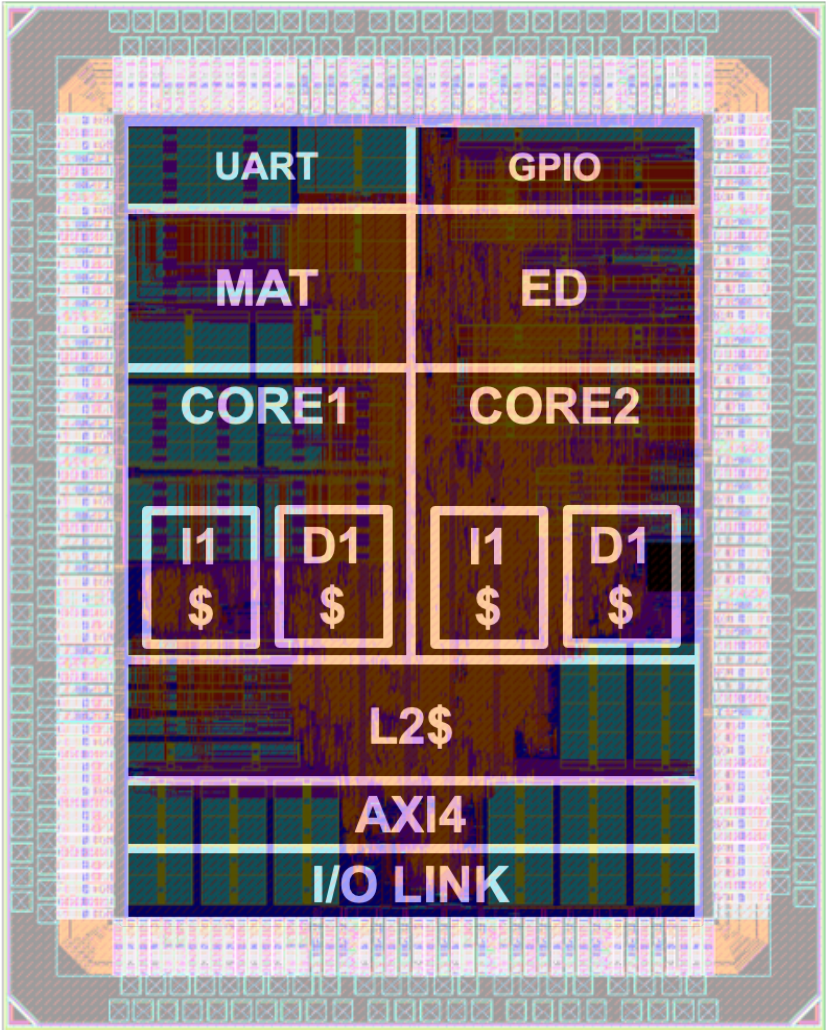}}
\caption{A 22-nm SoC for mobile genomics processing at the edge.}
\label{f:soc}
\end{figure}

The use-case and algorithm landscape is vast, so a discussion of this length can only hint at narrow fraction of the possibilities.  To that end we present the SoC design outlined in Fig.~\ref{f:soc}.  The chip consists of two 64-b Linux-capable in-order RISC-V cores (CORE1 and CORE2) with floating-point units (FPUs) and two bioinformatics accelerators, a 4${\times}$4 systolic array for matrix math (MAT) and an edit distance engine (ED).  The chip also has 700-KB on-silicon SRAM, split across the cache and accelerator needs.  Including I/O pads, the entire design consumes 5\,mm$^2$ in GlobalFoundries 22-nm CMOS FDSOI process.  The ability to realistically meet genomics needs with open-source RISC-V microarchitectures is a positive sign for the potential adoption of this approach by developers.

This design tries to strike a balance between flexibility and performance.  The two microprocessors clearly provide this flexibility by being able to run generic workloads, while the accelerators provide hardware speed-up for kernels fundamental to interesting functions in modern genomics pipelines.  In particular, CORE1 and CORE2 can facilitate many of the small intermediate support processes needed for practical implementation of a sequencing pipeline.  This includes steps like demultiplexing reads by their sample of origin (a low-cost comparison un-gapped string comparison), basic editing (primer trimming, chunking), filtering, and normalization.  Retaining FPUs in each core assists with the larger computational needs of such work which can be carried out in parallel with accelerator jobs if sufficient scratchpad memories are committed to MAD and ED.

The accelerators address the most fundamental front-line computations outlined above: DL basecalling and DP sequence comparison.  Together, along with the general computing ability of CORE1 and CORE2 can serve as an engine for rapid pathogen detection: the basecaller converting raw data to reads with the help of MAT, and ED quickly comparing it to some sample of a pathogenic genome.  In the case of viruses where many pandemic causing viruses have genomes below 30K\,bases in length, the opportunity to house sufficient computing within a Mobile-tier platform, perhaps even a Tiny-tier device, is good.

In this case we decided to take maximum advantage of our matrix-matrix multiplication engine by implementing a purely CNN-based basecaller.  Our design consists of six layers separated by ReLU activations and requires about 450K\,parameters in total. About 80\% of the weights reside in two layers, and very roughly, the basecaller is designed to deconvolve the contributions of raw signals over a window of 8 bases.  The final accuracy is 85\% which is insufficient for in-depth clinical applications, but practical for targeted pathogen  detection.

While computing under the control of a Linux system the SoC consumes about 50-mW peak at 250-MHz.  Accelerated basecaller performance is about 15${\times}$ faster and 13${\times}$ more energy efficient compared to core-only execution.  However, the absolute rate of basecalling at these rates still falls short of the real-time measurement capabilities of DNA sequencers and would require buffering or possibly pre-filtering to match measurement throughput.  Likewise a boost in clock rate and expansion of the MAT accelerator could help the performance.

Unfortunately, a Linux-induced communications bug introduced a deadlock in our CORE2-to-ED communications preventing us from being able to test ED on chip.  But FPGA base d results demonstrate the ability of this block to carry out DP comparisons between 100-base sequences at rates about 40${\times}$ faster than core-only runs and the ability to compare about 900K\,bases per second under a 250-MHz clock.

\section{Discussion \& Conclusion}

Edge hardware, is defined in this paper in terms of the components immediately bordering the RAN.  Such devices offer many exciting opportunities for genomics, especially for hosting the AI kernels needed by emerging mobile genomics applications.  In this paper we sought to outline and quantify these opportunities.  We used the MLCommons (MLC) benchmarking suite as a guide to better formalize problem size currently expected of hardware across the network's edge.  As a result, we considered devices partitioned across MLC's Tiny, Mobile, and Edge tiers with the ability to host deep learning models ranging across roughly 10K-1B parameters.  In MLC, these tiers cover deep learning models applied to, among others, keyword detection, automatic speech recognition, image segmentation, and natural language processing.  All of these jobs are highly relevant to genomics workloads which naturally require sequence analysis.  Size-wise, many relevant deep learning genomics kernels also match MLC's Edge/Mobile/Tiny benchmarks.

Only a minuscule number of fabricated digital CMOS ASICs focused on improving mobile sequencing at the edge have actually been described in the open literature.  In this paper we highlight three of these which address a critical component of sequencing pipelines: sequence comparison using fast indexing and dynamic programming.  These appear throughout the many genomics applications and so are clearly essential, but none of them address genomic AI in terms of deep learning.  To address this, this paper outlines a fabricated 22-nm SoC that attempts to do so.  The design combines matrix and sequence comparison accelerators with RISC-V cores and is targeted for the basecalling and lightweight alignment needs as may be appropriate for pathogen detection use cases.  The system can operate within a 50-mW power window and 250-MHz clock and thus holds potential for applications in the Tiny space.  

\bibliographystyle{IEEEtran}
\bibliography{mwscas25}

\end{document}